\documentclass[]{emulateapj}
\usepackage{graphics,amsmath}
\begin{document}

\title{Destruction of Binary Minor Planets During Neptune Scattering}
\author{\bf Alex H. Parker$^1$ and JJ Kavelaars$^2$}
\email{alexhp@uvic.ca}
\affil{\emph{ $^{1}$Department of Astronomy, University of Victoria,  $^{2}$Herzberg Institute of Astrophysics, National Research Council of Canada}}

\shortauthors{Parker \& Kavelaars}

\begin{abstract}

The existence of extremely wide binaries in the low-inclination component of the Kuiper Belt provides a unique handle on the dynamical history of this population. Some popular frameworks of the formation of the Kuiper Belt suggest that planetesimals were moved there from lower semi-major axis orbits by scattering encounters with Neptune. We test the effects such events would have on binary systems, and find that wide binaries are efficiently destroyed by the kinds of scattering events required to create the Kuiper Belt with this mechanism. This indicates that a binary-bearing component of the cold Kuiper Belt was emplaced through a gentler mechanism or was formed \emph{in situ}.

\end{abstract}


\maketitle

\section{Introduction}

The dynamical histories of the sub-populations of the Kuiper Belt remain in contest. Producing an inclination width as high as is observed (eg., Kavelaars et al. 2008) while maintaining a cold component with low eccentricities and inclinations remains a theoretically challenging prospect. Levison et al. (2008, hereafter L08) present numerical simulations designed to determine if the Nice model, a proposed framework for understanding the history and architecture of the outer Solar System, provides a pathway to create the low-inclination component of the classical Kuiper Belt. They suggest that the objects that now reside in the Cold Classical Kuiper Belt (CCKB) may have formed significantly interior to their present heliocentric distances, and were scattered outward by interactions with Neptune during a high-eccentricity period of its early orbital evolution. During this period, the region exterior to Neptune was full of overlapping mean-motion resonances (MMRs). Objects that entered this region could chaotically evolve to low eccentricity, where some would become trapped as the eccentricity of Neptune damped out and the resonance widths decreased and ceased to overlap. Assuming that Neptune's eccentricity started at 0.3 and was damped with a timescale of order 1 Myr, L08 finds that a distinct population of low-$e$, low-$i$ objects can be injected into the Kuiper Belt region. 

Trans-Neptunian Binaries (for review see Noll et al. 2008a) likely formed in the primordial disk (binary formation in the current Kuiper Belt is suppressed by its low density and relatively high energetics) and would have been subjected to the same Neptune interactions proposed for the entire CCKB-precursor population by L08. The current binary population in the CCKB has a component with extremely wide separations; for example, the system 2001 QW$_{322}$ has a semi-major axis to Hill radius fraction ($a/R_{H}$) of $\sim0.27-0.32$ (Petit et al. 2008). Such wide systems are extremely sensitive to perturbations (Petit and Mousis, 2004), and may have undergone significant modification or even disruption if they were subjected to close encounters with Neptune. We seek to test whether an initial population of TNBs subjected to emplacement into the CCKB through the L08 mechanism could leave behind a remnant population of binaries large enough to produce those we see today. We track the evolution of a given binary system through a suite of \emph{multiple} of close encounters with Neptune determined by the close encounter histories of objects emplaced in the CCKB through the L08 mechanism.

Binary-planet encounters have been explored in the context of capturing planetary satellites (Agnor and Hamilton, 2006; Vokrouhlick\'y et al. 2008). Our analysis and simulations are similar to those used by these studies, but specifically focus on the end-states of the \emph{binaries}, rather than the captured population of planetary satellites.

\begin{table}
\centering
\footnotesize
\begin{tabular}{lc}
\multicolumn{2}{c}{\bf Table 1}\\
\multicolumn{2}{c}{\bf Initial Planetesimal Orbits}\\
\hline
\hline
Parameter & Range \\
\hline
$a_{\mbox{out}}$ & 20--34 AU \\
$e_{\mbox{out}}$ & 0--0.4 ($a_{\mbox{out}}<29$ AU), 0--0.15 ($a_{\mbox{out}}>29$ AU) \\
$i_{\mbox{out}}$ & 0$^\circ$--10$^\circ$ ($a_{\mbox{out}}<29$ AU), 0$^\circ$ ($a_{\mbox{out}}>29$ AU)\\
\hline
 \end{tabular}\\
 \end{table}

\begin{figure*}
\begin{centering}
\includegraphics[width=15cm]{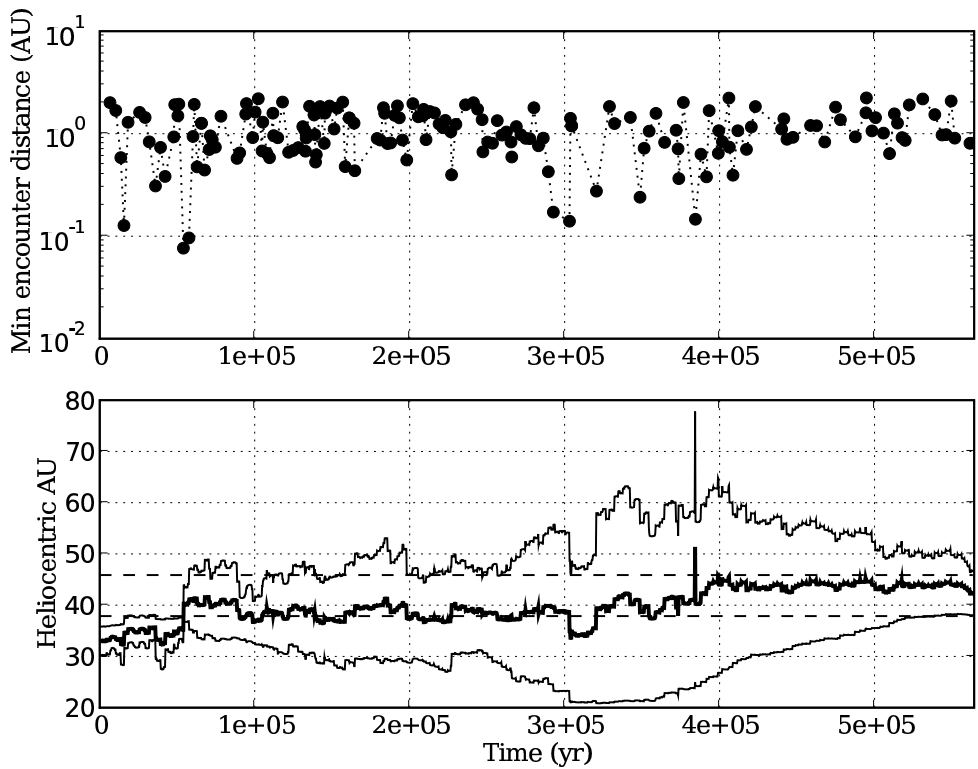}
\caption{An example of a close encounter history for one particle flagged as having reached the CCKB. Top Panel: minimum Neptuneocentric distance for all tracked close encounters vs. time since integration start. Bottom panel: Orbital elements of the particle vs. time since integration start. Thick line represents semi-major axis, while lower and upper thin lines represent pericenter and apocenter, respectively. Lower and upper dashed lines mark our synthetic Neptune's 3:2 and 2:1 mean-motion resonances, respectively. Only close encounters inside $1 \mbox{AU}$ were treated during our integrations of binary mutual orbits.}
\end{centering}
\end{figure*}

\section{Separations of Wide Binaries}

The characteristic ratio we used to describe the ``wideness'' of a binary is the mutual semi-major axis to Hill Radius ($a/R_H$), and can be defined as follows:

\begin{equation}
a/R_H = a_{\mbox{out}} \times \left( \frac{ 3 M_{\odot}}{ M_p + M_s } \right)^\frac{1}{3} ,
\end{equation}

\noindent where $a_{\mbox{out}}$ is the heliocentric semi-major axis, and $M_p$ and $M_s$ are the masses of the primary and secondary components of the binary, respectively. We can convert $a/R_H$ to another commonly used ratio, the mutual semi-major axis to the ``Effective Radius'' of the system ($a/R_E$), where 

\begin{equation}
R_E = \left( R_p^3 + R_s^3 \right)^\frac{1}{3} ,
\end{equation}

\noindent and $R_p$ and $R_s$ are the radii of the primary and secondary components of the binary. To convert between the two ratios, we use the following relationship:

\begin{equation}
a/R_E = ( a/R_H ) \times a_{\mbox{out}} \left( \frac{4\pi\rho}{9M_\odot} \right)^\frac{1}{3}.
\end{equation}

For all conversions in this work, we assumed $\rho = 1$ g cm$^3$.

For the system 2001 QW$_{322}$, mutual orbital elements have been determined after several years of monitoring (Petit et al. 2008). For most other binary systems with wide discovery separations, only the discovery epoch data is in the literature. Grundy et al. (2008) show that the most common on-sky separation of a binary is  $0.99 \times a$, allowing us to estimate the semi-major axis of a system based on its discovery separation. We can also estimate a mass for each component by assuming an albedo and density for the system, given photometry for the components. We assume an albedo of 0.16, the value adopted by Petit et al. (2008) for 2001 QW$_{322}$ and consistent with the high albedos measured for Cold Classical objects (eg., Brucker et al. 2009).

Using discovery separation and photometry, we estimate $a/R_E$ for three additional systems: 2000 CF$_{105}$ (IAUC 7857), 2003 UN$_{284}$ (IAUC 8251), and 2005 EO$_{304}$ (IAUC 8526)\footnote{$\Delta$-mag taken from original observations, magnitude of system taken from the Minor Planet Center --- http://www.cfa.harvard.edu/iau/mpc.html}. Distances at discovery were extracted using the Minor Planet Center ephemerides. Given these measurements, for these three systems we estimate $a/R_E$ to be approximately 531, 979, and 1017, respectively. These values will be compared to the results of our numerical integrations of Neptune encounters, which are presented in the following sections.

\section{Close Encounter Histories}

To accurately determine the tidal effects of Neptune on passing binaries, we first determined the trajectories of their close approaches relative to Neptune. These trajectories were generated by running a suite of N-body simulations that track the interactions of a Neptune-mass object with a swarm of massless particles. Since we are interested in testing the effects of the Kuiper Belt formation scenario simulated by L08, we must generate close-encounter trajectories that reflect those that binaries would have experienced given the L08 model. The initial conditions of our simulations were chosen to emulate Run B from L08, since this run was the most successful at generating a final Kuiper Belt with characteristics similar to those we see today. This initial configuration sets Neptune's initial semi-major axis at 28.9 AU and eccentricity at 0.3, and the swarm of planetesimals initially between 20 AU $< a_{\mbox{out}} <$ 34 AU (see Table 1 for initial orbital elements of the planetesimal swarm).

Since we are interested only in the characteristics of the close encounters that scatter objects into orbits that can be caught in the series of overlapping exterior MMRs, we did not evolve Neptune's orbital elements from their initial values. We monitored each particle's semi-major axis and eccentricity for 1 Myr and find all those candidates that pass through ``CCKB-like'' orbits which no longer have close encounters with Neptune. If any particle passed through 38 AU $< a_{\mbox{out}} <$ 45.9 AU (3:2 and 2:1 MMRs for the initial Neptune orbit), $e_{\mbox{out}} < 0.1$, and $i _{\mbox{out}}< 5^\circ$ after at least 0.5 Myr has passed, we flagged it as having reached the CCKB. This is a conservative approach, as objects may continue to evolve in eccentricity for some time before the overlapping MMRs would have snapped closed due to the eccentricity damping imposed on Neptune by L08, which could result in later epochs where the object has close encounters with Neptune. Additionally, by allowing such low (38 AU) semi-major axes for our CCKB flagged sample, we are including objects that did not suffer as much orbital migration as would be required for objects reaching the real CCKB, which exists between semi-major axis limits of $\sim42.5-47$ AU (eg., Kavelaars et al. 2008).

\begin{figure}[t]
\begin{centering}
\includegraphics[width=9cm]{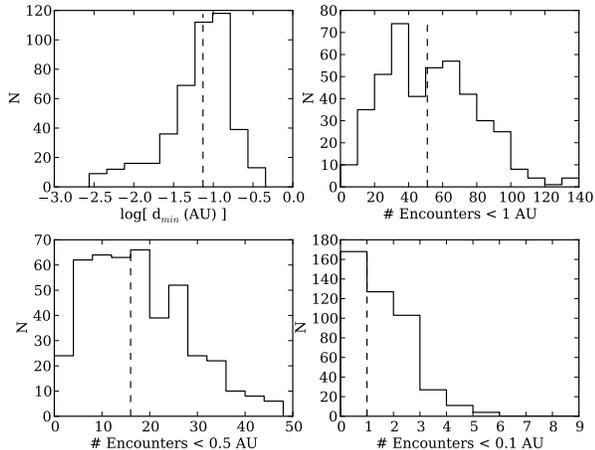}
\caption{Close encounter statistics for all 156 particles which were flagged as entering the CCKB. Top left: histogram of the minimum separation for the closest approach in each close encounter history. Top right, lower left, and lower right: Histogram of the number of encounters in each close encounter history $<$ 1 AU, 0.5 AU, and 0.1 AU, respectively. In all panels, medians are illustrated with vertical dashed lines. For conversions, Neptune's hill radius is approximately 0.77 AU, and 0.1 AU correspond to approximately 600 Neptune radii.}
\end{centering}
\end{figure}

Using a modified version of the \emph{Mercury 6} N-body code (Chambers 1999), we integrated an initial population of 15,000 particles; of these, 156 were emplaced in the CCKB according to our criteria. For every object which we flagged as having reached the CCKB, we generated a ``close encounter history'' by extracting all encounters that occurred within 3 Neptune Hill Radii. Figure 1 illustrates a single close encounter history for a particle which reaches the CCKB in $6\times10^5$ years, and Figure 2 shows the distribution of close encounters for all particles flagged as reaching the CCKB in our simulations. 

\begin{figure}[t]
\begin{centering}
\includegraphics[width=9cm]{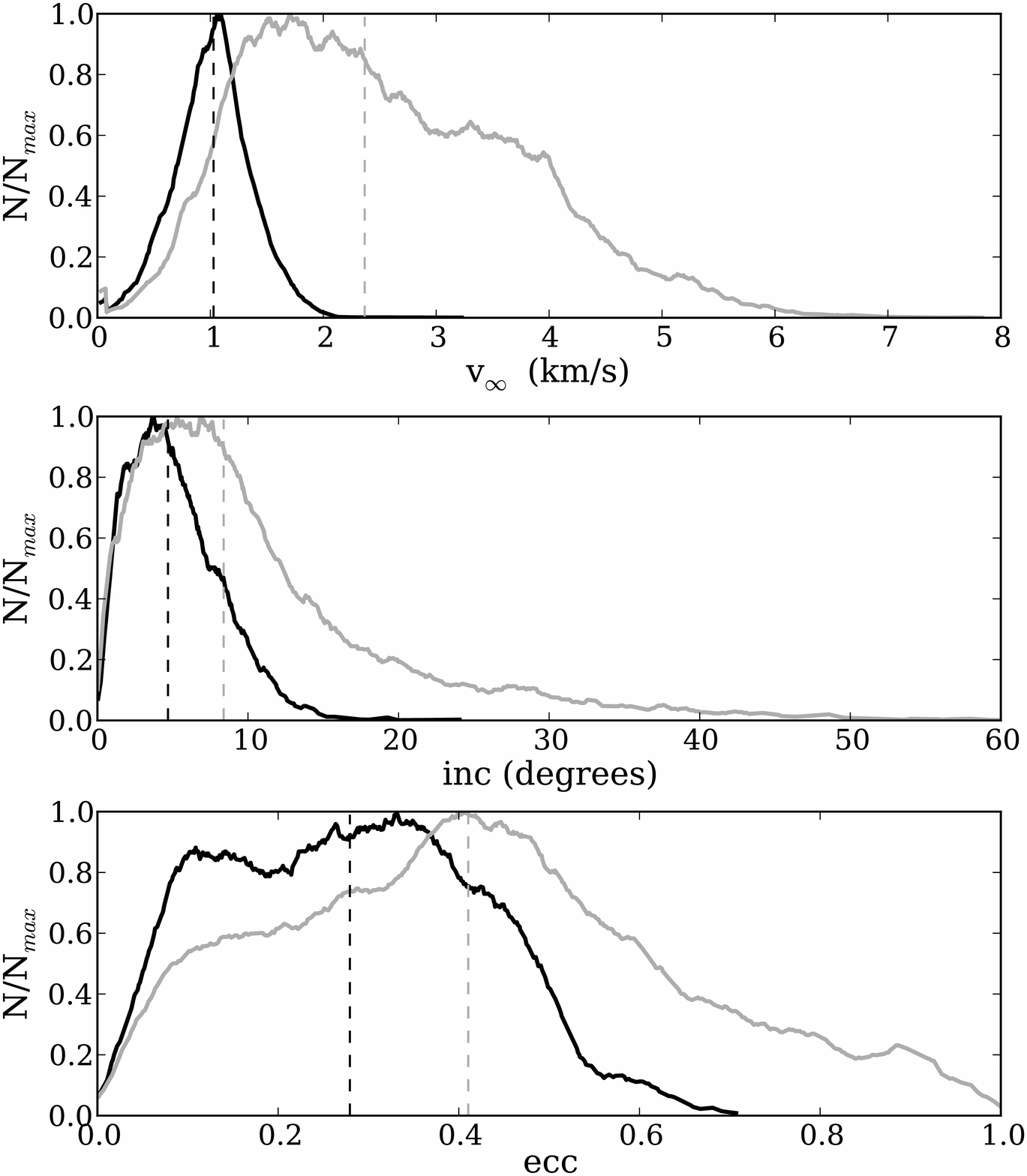}
\caption{Top panel: Distribution of velocity at infinity (relative to Neptune) for close encounters in our simulations, normalized to unity. Middle and bottom panel: Distribution of inclination and eccentricity for particles in our N-body simulations. Values are sampled at random epochs from all particles' histories. Black lines represent the distribution for particles that reached CCKB. Gray lines represent the distributions for random subset of all particles. Vertical dashed lines mark medians of each distribution.}
\end{centering}
\end{figure}

The top panel of Fig. 3 illustrates the relative velocity at infinity $v_{\infty}$ for the Neptune close encounters in our simulations. The distribution for the particles that reached the CCKB is peaked at approximately 1 km s$^{-1}$, and the peak is coincident with the median. The peak for the distribution of all the particles is at approximately 1.7 km s$^{-1}$, while the median lies at 2.3 km s$^{-1}$. This is consistent with the distributions produced by the simulations presented in Vokrouhlick\'y et al. (2008). The bottom two panels of Fig. 3 show the source of the difference in the $v_{\infty}$ distributions between the total population of particles and the subset that reach the CCKB. The crossing velocity between two objects is driven by their relative inclination and eccentricity, and objects that reach the CCKB have a significantly lower median inclination and eccentricity throughout their history. In other words, objects that \emph{start with} or that are \emph{scattered into} highly inclined or highly eccentric orbits are rarely emplaced into the CCKB. Because of this, the median $v_{\infty}$ for the particles that reach the CCKB is lower than for the entire population (which includes many objects scattered to large inclination or eccentricity).

\section{Integration of Binary Orbits}

We used a second, purpose-built numerical integrator to simulate the close encounters between binary systems and Neptune. \emph{Mercury 6} was modified to produce position and velocity at closest approach to Neptune for each close encounter passing less than 1 AU ($\sim1.3$ Neptune Hill radii) from Neptune. For each close encounter, we generated a hyperbolic trajectory with respect to Neptune that replicated these quantities (Neptuneocentric pericenter separation and velocity). If a given encounter lead to temporary capture by Neptune, we increased the encounter velocity to the parabolic limit, decreasing the overall tidal impulse Neptune can exert on the system (making this a conservative approximation with respect to the efficiency of disruption of binaries). 

All binaries are treated as a massive primary with a test particle as a secondary.  We verified that this had little effect on our results by running a sample of several thousand binary integrations with mass ratios of 1:1 which produced no notable change in outcome.

The initial binary orbits are sampled from the range of covered by known classical Kuiper Belt binaries (Naoz et al. 2010, Grundy et al. 2009, Petit et al. 2008, Veillet et al. 2002), and are generated as follows: 
\begin{enumerate} 
\item Draw $a/R_H$ from a uniform distribution between 0.005---0.3.
\item Draw $a$ from from a uniform distribution between $2\times10^3$---$1.2\times10^5$ km.
\item Determine $M$ from given $a/R_H$ and $a$. If it lies outside $10^{16}$---$10^{19}$ kg, re-select $a$ and re-compute $M$. Repeat until successful.
\item Eccentricity is drawn from a uniform distribution between 0---0.95, and all initial angular parameters (including inclination) are drawn from uniform distributions over their entire physical ranges.
\end{enumerate}

At the beginning of each integration of a given close encounter in a specific encounter history, we used the mutual orbital elements from the \emph{last} integration of the same binary in order to generate the relative position and velocity of the secondary with respect to the primary (mean anomaly, node, and argument of pericenter are randomized). We then placed the primary on the hyperbolic trajectory of this close approach with Neptune, and placed the secondary on the same trajectory with the addition of an offset in position and velocity determined by its mutual orbit with the primary. The binary is started at a distance of 1.5 AU ($\sim2$ Neptune Hill radii) from Neptune, and integrated until passing outside a radius of 1.5 AU from Neptune. All motions are taken to be in the frame of Neptune.

A binary is considered destroyed if any of the following criteria are met after any close encounter integration: 

\begin{enumerate}
\item The system becomes unbound (Total energy of the system is greater than zero).
\item Mutual semi-major axis is enlarged to more than one Hill Radius (with respect to the Sun, assuming a circular final heliocentric orbit at 40 AU). 
\item Binary components collide (assuming a density for each component of 1 g cm$^3$ in order to determine radii).
\end{enumerate}

 At the end of \emph{all} close encounters in a given history, a binary is considered to have survived emplacement into the CCKB if it has not been destroyed by any of the criteria above. In our simulations, by far the most common mechanism for destruction was (2), representing $\sim$98.5\% of all disruptions, while cases (3) and (1) each contributed nearly negligibly at $\sim$1.3\% and $\sim$0.2\%, respectively. In most cases, the closest few approaches in a given encounter history determine if a binary survives (mutual semi-major axis evolution is similar to a L\'{e}vy flight, eg., Colins and Sari 2008); however, more distant encounters can cause modification of binary orbits that make later close approaches more likely to disrupt the system. This illustrates the importance of treating entire encounter histories.

These destruction criteria should be considered very conservative. The true $a/R_H$ limit for stability varies with inclination but is always $<1$, and considering $R_H$ variations due to heliocentric orbit evolution during scattering will lead to more efficient disruption by solar tides. Mergers may become more important if we considered Kozai cycles (Perets and Naoz 2009, Kozai 1962) and tidal damping effects.

\begin{figure*}[t]
\begin{centering}
\includegraphics[width=15cm]{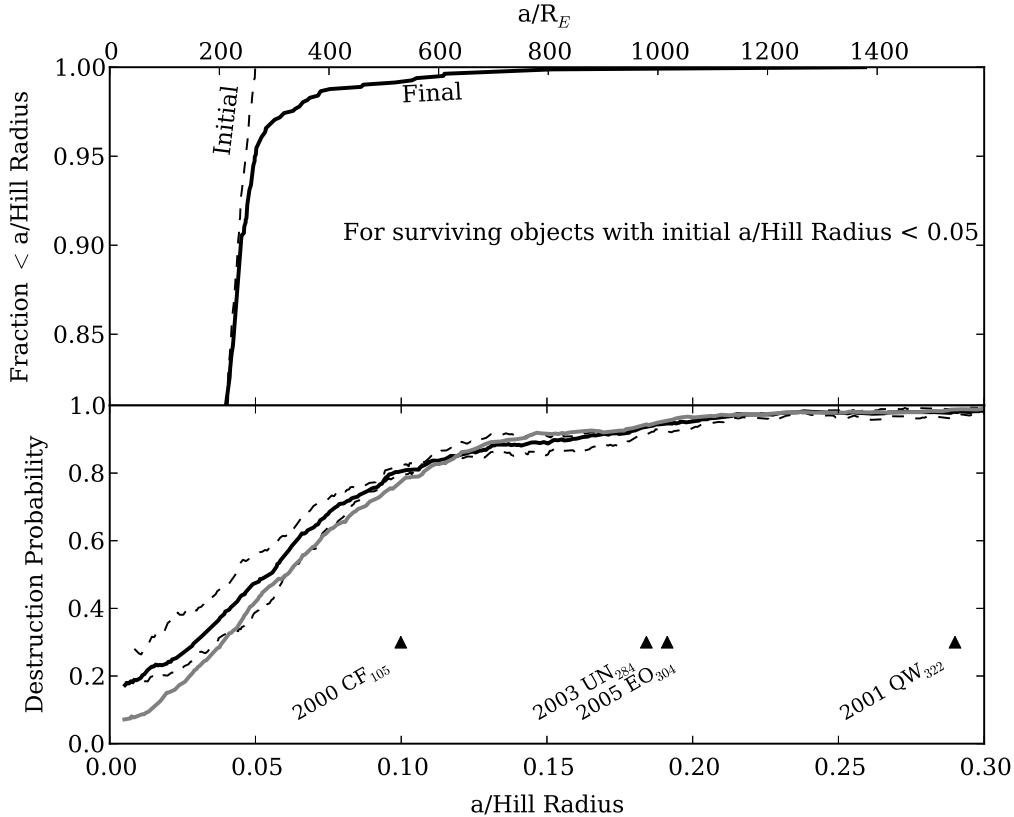}
\caption{Results from two sets of 7,500 binary-Neptune integrations. Top panel: ``Mobility'' of initially tight binaries. Dashed line is a cumulative histogram of $a/R_H$ (or $a/R_E$ on top axis) prior to Neptune interactions for surviving binaries with initial $a/R_H < 0.05$. Solid line is a cumulative histogram of $a/R_H$ for the same binaries after interactions. Bottom panel: Probability of destruction of a binary system as a function of its initial $a/R_H$. Lower and upper dashed lines represent subset of sample with $e<0.2$ and $e>0.7$, respectively. Gray line: Results from integrating encounter histories for objects with initial $a_{\mbox{out}}>29 \mbox{AU}$. Triangles: estimates of $a/R_E$ for known wide binaries.}
\end{centering}
\end{figure*}

Figure 4 illustrates the results of 7,500 integrations of the binary-Neptune interactions, requiring roughly 150 CPU-hours to integrate. Binaries with initially wide orbits are efficiently destroyed before being emplaced in the CCKB, with the probability of destruction crossing 50\% at approximately $a/R_H\sim0.05$ and increasing to 80\% by $a/R_H\sim0.1$. Additionally, we find that it is unlikely that binaries with initially tight orbits and which survive emplacement are left with significantly widened orbits: for all binaries that survived emplacement with initial $a/R_H<0.05$, only $\sim5\%$ have final $a/R_H$ exceeding 0.05, and fewer than 2\% exceed 0.075. Those binaries that are widened show no preference for any specific region of initial mutual eccentricity-inclination phase space.

We found a larger fraction ($\sim50$\%) of objects implanted into the CCKB from inside the 29 AU boundary between the initially excited and non-excited disks than was found by L08. To check if these additional inner objects skewed the probability of binary disruption (as L08 showed that such objects are subjected to more Neptune encounters), we performed a second set of 7,500 binary encounter integrations using only the 94 encounter histories for objects with initial $a_{\mbox{out}} > 29 \mbox{AU}$, removing the influence of the inner objects. The results are illustrated in Figure 4, and for wide separations the variations in destruction probability between the full sample and the ``outer disk'' sample are minor.

Binary mutual orbits that were initially very eccentric ($e>0.7$) were somewhat more easily disrupted than initially more circular orbits ($e<0.2$) for moderate $a/R_H$, with the largest variation between the two groups being $\sim10$\%. For higher $a/R_H$, this difference decreases substantially. A second smaller set of integrations was run to test the effect of initial inclination, and the difference in destruction probability between separate samples of 1500 $i=0^\circ$ and $i=90^\circ$ mutual orbits remained less than a few percent at all $a/R_H$.

\section{Discussion}

We have simulated the effects of interactions between Neptune and Trans-Neptunian Binaries, given the kinds of encounter histories required for emplacement of the CCKB though the mechanism proposed by L08, and we find that wide binaries ($a/R_H \gtrsim 0.05$) are efficiently destroyed by these interactions. Any primordial population of wide binaries would be decimated by emplacement into the cold belt through this process. 

If a population of primordial wide binaries with $a/R_H$ similar to the four illustrated in Figure 4 were subjected to these kinds of scattering events, an initial population of $\sim150$ would be required to leave the four survivors behind. Alternatively, if the binaries were initially tight ($a/R_H < 0.05$), then an initial population of $\sim300$ binaries would be required to leave four survivors with sufficiently widened orbits behind. Additionally, due the destruction of binaries by collisions with interloping Trans-Neptunian Objects over the age of the Solar System, a much (at least 10 times) larger primordial population of wide binaries may be required to produce the remaining population we see today (Petit and Mousis 2004). Together, these results would require more primordial binaries than even a binary fraction of 100\% would allow, suggesting that wide binaries such as 2001 QW$_{322}$ were not subjected to a period of close encounters with Neptune. This implies that a component of the Cold Classical Kuiper Belt was not emplaced by scattering from lower heliocentric orbits, but either had to form \emph{in situ} or be implanted by a gentler mechanism in order to preserve its population of wide binaries. A caveat is that the evolution of binary properties over the age of the Solar System is poorly understood, and collisions or other perturbations may cause primordially tight binaries to widen post-emplacement. If this is the case, however, we would expect wide binaries to exist in other Trans-Neptunian populations that have been subjected to similar long-term collisional evolution as the CCKB.

Binary destruction by Neptune scattering may help explain the variation of binary fraction between different Trans-Neptunian populations.  The binary fraction is much lower in the high-$i$ (Hot) component of the Classical Kuiper Belt than in the low-$i$ component (Noll et al. 2008b). L08 illustrate that objects captured in the Hot Classical Kuiper Belt via the mechanisms presented in Gomes (2003) interacted with Neptune for significantly longer than simulated in this work ($\sim10^7$ years). If the Hot Classical Kuiper Belt was emplaced in large part by these mechanisms, a significant fraction of its initial binary population may have been destroyed. This would naturally decrease the binary fraction of this population without appealing to variations in the original formation mechanism of Trans-Neptunian Binaries. 

The orbital properties of known binaries presented in this work (excluding 2001 QW$_{322}$) are based on a number of assumptions, and should be regarded with caution. We have obtained full astrometric orbital solutions for all binaries discussed here, and our results remain unchanged using the new, accurate values. These orbital solutions will be presented in an upcoming paper.

The two known binary Centaurs (Grundy et al. 2008; Grundy et al. 2007) are actively interacting with the giant planets, and have dynamical lifetimes of order 1-100 Myr (Tiscareno and Malhotra 2003). If we take our simulations of encounters with Neptune as a baseline for estimating the probability of disruption of these binary systems over their lifetimes as Centaurs, we see that for systems as tight as these (with $a/R_E$ of order $\sim$20), probability of disruption after $\sim1$ Myr is fairly low (15-20\%), but their survival over tens of Myr is quite unlikely. Their existence likely reflects the different encounter histories Centaurs are subjected to compared to objects injected into the CCKB through the mechanisms studied here.

\clearpage

\section{Acknowledgements}

Alex H. Parker is supported by the NSF-GRFP award DGE-0836694. This research used the facilities of the Canadian Astronomy Data Centre operated by the National Research Council of Canada with the support of the Canadian Space Agency. We acknowledge the many useful discussions with Wesley Fraser, Brett Gladman, and Sarah Greenstreet, and additionally wish to thank Stephen Gwyn and John Ouellette for their assistance with the computational challenges of this work.


\end{document}